\documentclass{phbauth}
\usepackage{graphicx}
\begin{document}
\begin{frontmatter}
\title{Quantum Phase Interference in Magnetic Molecular Clusters}

\author[address1]{W. Wernsdorfer\thanksref{thank1}},
\author[address1]{I. Chiorescu}
\author[address2]{R. Sessoli}
\author[address2]{D. Gatteschi}
\author[address3]{D. Mailly}

\address[address1]{Lab. L. N\'eel - CNRS, BP166, 38042 Grenoble, France}
\address[address2]{Dept. of Chem., Univ. of Florence, 50144 Firenze, Italy}
\address[address3]{LMM - CNRS, 196 av. H. Ravera, 92220 Bagneux, France}

\thanks[thank1]{Corresponding author. E-mail: wernsdor@labs.polycnrs-gre.fr} 

\begin{abstract}
The Landau Zener model has recently been used to measure very 
small tunnel splittings in molecular clusters of Fe$_8$, which at low temperature behaves like 
a nanomagnet with a spin ground state of $S$~= 10. The observed oscillations of the 
tunnel splittings as a function of the magnetic field applied along the hard anisotropy axis 
are due to topological quantum interference of two tunnel paths of opposite windings. 
Transitions between quantum numbers $M = -S$ and $(S - n)$, with $n$ even or odd, 
revealed a parity effect which is analogous to the suppression of tunnelling predicted for 
half integer spins. This observation is the first direct evidence of the topological part of the 
quantum spin phase (Berry or Haldane phase) in a magnetic system. We show here that the quantum interference can also be measured by ac susceptibility measurements in the thermal activated regime.
\end{abstract}

\end{frontmatter}


Studying the limits between classical and quantum physics has become a very attractive 
field of research which is known as 'mesoscopic' physics. New and fascinating 
mesoscopic effects can occur when characteristic system dimensions are smaller than the 
length over which the quantum wave function of a physical quantity remains sensitive to 
phase changes. Quantum interference effects in mesoscopic systems have, until now, 
involved phase interference between paths of particles moving in real space as in SQUIDs 
or mesoscopic rings. For magnetic systems, similar effects have been proposed for 
spins moving in spin space, such as magnetisation tunnelling out of a metastable potential 
well, or coherent tunnelling between classically degenerate directions of magnetisation \cite{Chichilianne94}. 
Up to now, magnetic molecular clusters have been the most promising candidates to 
observe these phenomena since they have a well defined structure with well characterised 
spin ground state and magnetic anisotropy. These molecules are regularly assembled in 
large crystals where often all molecules have the same orientation. Hence, macroscopic 
measurements can give direct access to single molecule properties. The most prominent 
examples are Mn$_{12}$ acetate \cite{Sessoli93} and Fe$_8$ \cite{Barra96} having a spin ground state of $S = 10$, and an Ising-type magneto-crystalline anisotropy.

The relaxation time of Fe$_8$ becomes temperature independent below 360 mK \cite{Sangregorio97} showing that a pure tunnelling mechanism between the only populated $M = \pm10$ states is responsible for 
the relaxation of the magnetisation. Fe$_8$ allowed us to observe directly the existence of what in a semi-
classical description is the quantum spin phase (Berry - Haldane phase \cite{Berry84,Haldane88}) associated with the magnetic spin of the cluster \cite{Wernsdorfer99}. The topological part of the phase leads to constructive (S integer) or destructive interference (S half-integer) between 
spin paths of opposite windings \cite{Loss92,Delft92} which can be directly evidenced by measuring the 
tunnelling splitting $\Delta$ as a function of a magnetic field applied along the hard anisotropy 
axis which was first proposed in \cite{Garg93}. Furthermore, we observed the predicted parity (symmetry) effects when comparing the 
transitions between different energy levels of the system \cite{Wernsdorfer99} which is analogous to the parity 
effect between systems with half integer or integer spins \cite{Loss92}.

\begin{figure}[t]
\begin{center}\leavevmode
\includegraphics[width=0.8\linewidth]{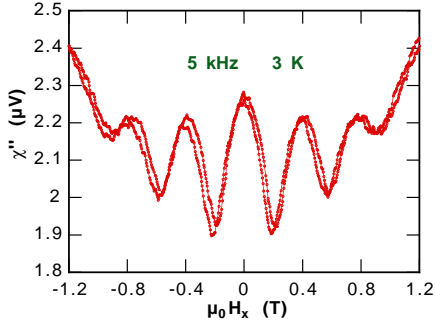}
\caption{Measured ac susceptibility $\chi''$ as a function of transverse field in direction of the hard axis $x$.}
\label{fig1}
\end{center}
\end{figure}

We show here that the quantum interference can also be measured by ac susceptibility measurements in the thermal activated regime. We used a home built Hall probe magnetometer, allowing to measure the ac susceptibility of single crystals on the order of 10 to 50 $\mu$m. It works in the temperature range between 30 mK and 30 K, for frequencies between 1 Hz and 100 kHz, and for applied fields up to 1.4 T. The field can be applied in any direction with a precision better than 0.1$^{\circ}$. Fig. 1 shows a typical measurement of ac susceptibility $\chi''$ as a function of transverse field in direction of the hard axis $x$. A clear oscillation of $\chi''$ can be seen which corresponds to an oscillation of the relaxation rate.

The simplest model describing the spin system of Fe8 molecular clusters is
the giant spin model with a spin $S = 10$. Using the same anisotropy constants as in Ref. \cite{Wernsdorfer99}, we calculated the tunnel splitting $\Delta$ as a function of a transverse field in direction of the hard axis $x$, for quantum transition between $M =  \pm10$, $\pm9$ and $\pm8$ (fig. 2). Note the decrease of the oscillation period for exited levels. This feature has been found experimentally (cf. Fig. 1 and Ref. \cite{Wernsdorfer99}).

\begin{figure}[t]
\begin{center}\leavevmode
\includegraphics[width=0.8\linewidth]{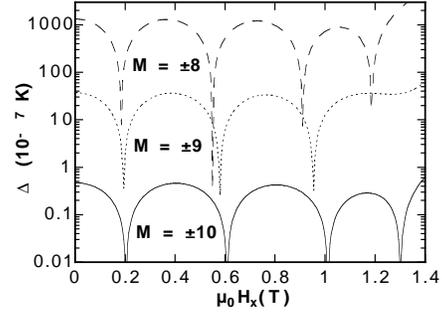}
\caption{Calculated tunnel splitting $\Delta$ as a function of transverse field in direction of the hard axis $x$ for quantum transition between $M =  \pm10$, $\pm9$ and $\pm8$.}
\label{fig2}
\end{center}
\end{figure}

%

\end{document}